# Solar Cycle 24: is the peak coming?

Stefano Sello
Mathematical and Physical Models
Enel Research
Via Andrea Pisano, N.120
56123 Pisa, Italy

Abstract

Solar cycle activity forecasting, mainly its magnitude and timing, is an essential issue for numerous scientific and technological applications: in fact, during an active solar period, many strong eruptions occur on the Sun with increasing frequency, such as flares, coronal mass ejections, high velocity solar wind photons and particles, which can severely affect the Earth's ionosphere and the geomagnetic field, with impacts on the low atmosphere. Thus it is very important to develop reliable solar cycle prediction methods for the incoming solar activity. The current solar cycle 24 appeared unusual from many points of view: an unusually extended minimum period, and a global low activity compared to those of the previous three or four cycles. Currently, there are many different evidences that the peak in the northern hemisphere already occurred at 2011.6 but not yet in the southern hemisphere. In this brief note we update the peak prediction and its timing, based on the most recent observations.

It is well known that it is quite reliable a forecasting of the maximum of a solar cycle using the coronal Fe XIV emissions (see: Altrock, 2003; 2012). In particular, the data show that solar maximum in previous three cycles 21-23, occurred when the center line of the "Rush to the Poles" of polar crown prominences reached a critical latitude of 76° ± 2°. Furthermore, we see that in the previous three cycles, solar maximum occurred when the number of Fe XIV emission first reached latitudes 20° ± 1.7°. However, it is also well known the unusual nature of Cycle 24: indeed we detected an intermittent Rush that is only well-defined in the northern hemisphere. In particular, in 2009 an initial low slope of 4.6°year$^{-1}$ was found in the north, compared to an average of 9.4 ± 1.7 ° year$^{-1}$ in the previous cycles. Altrock showed that an early fit to the Rush would have reached the critical latitude of 76° at 2014.6. However, in 2010 the above slope increased to 7.5° year$^{-1}$ differently to the previous three cycles. On the other hand, there are clear evidences that the solar maximum, given by the monthly smoothed sunspot numbers, in the northern hemisphere already occurred at 2011.6 ± 0.3. In the southern hemisphere there is now an increase of the activity and this indicates solar maximum is occurring now in the north but not yet in the south. Altrock showed that the Rush to the Poles, if it exists, is very poorly defined. In fact, a linear fit to several maxima would reach 76° in the south at 2014.2. Currently the greatest number of emission regions is at 21° in the north and 24°in the south. This indicates that solar maximum is occurring now in the north but not yet in the south. For more details see: Altrock, 2012.

Gopalswamy et al. (2012) come to similar conclusions from a study of microwave brightness and prominence eruptions. In addition, there is also strong evidence that the northern-hemisphere sunspot *area* reached a maximum around the end of 2011. The southern-hemisphere area appears to still be increasing. We can clearly see an inflection point in the global monthly smoothed sunspot *number* late in 2011, (20011.6), which could be evidence that maximum occurred in one hemisphere.

The question now is: when will occur the solar cycle peak in correspondence to the maximum activity in the southern hemisphere?

Currently, there are many predictions on the occurrence of the solar cycle 24 maximum, using different methods, both precursor and non-precursor. Recently, in particular, we showed a nonlinear full-shape curve prediction after the onset of the new solar cycle 24, indicating an estimated principal peak amplitude for the monthly smoothed sunspot numbers of 82 ± 5, timing June 2012, and this further supports some early indications, both using precursor and non-precursor methods, of a "moderately" weak new solar cycle. It is well known now that current solar cycle 24 appeared unusual from many points of view: an unusually extended minimum period, and a global low activity compared to those of the previous three or four cycles.

Figure 1 summarizes the predictions, based on the SIDC observed monthly smoothed sunspot numbers, updated to October 2010. In the same figure it is also shown the most recent observed data from SIDC updated to September 2012. As we can see, at around 2011.6 there was a stop in the increase of solar activity, indicating the occurrence of the solar maximum in the northern hemisphere, as already noted. After that date, the solar activity increased again but with a slower rate and currently the most probable prediction of the last increasing phase, is that we will reach the peak of the solar cycle with the observed data up to December 2012, corresponding to a monthly smoothed number near the value 75. Note that all the predictions somewhat overestimated the increasing rate phase toward the maximum. We recall here that the monthly smoothed sunspot number is shifted in time relative to the monthly sunspot number by an amount of six months. Thus, in order to check the validity of the above peak prediction we have to wait the observed data of the monthly sunspot numbers up to December 2012. We further note that the current IPS forecast is quite consistent with our prediction (i.e. monthly smoothed peak of 90.2 at December 2012, see Figure 2), but there are also many other current predictions that indicate a longer time (around October 2013) needed to reach the peak of the current solar cycle (e.g. see: Uzal et al., 2012). Only the future observations (at least for the next four or five months) will decide which prediction was the most correct.

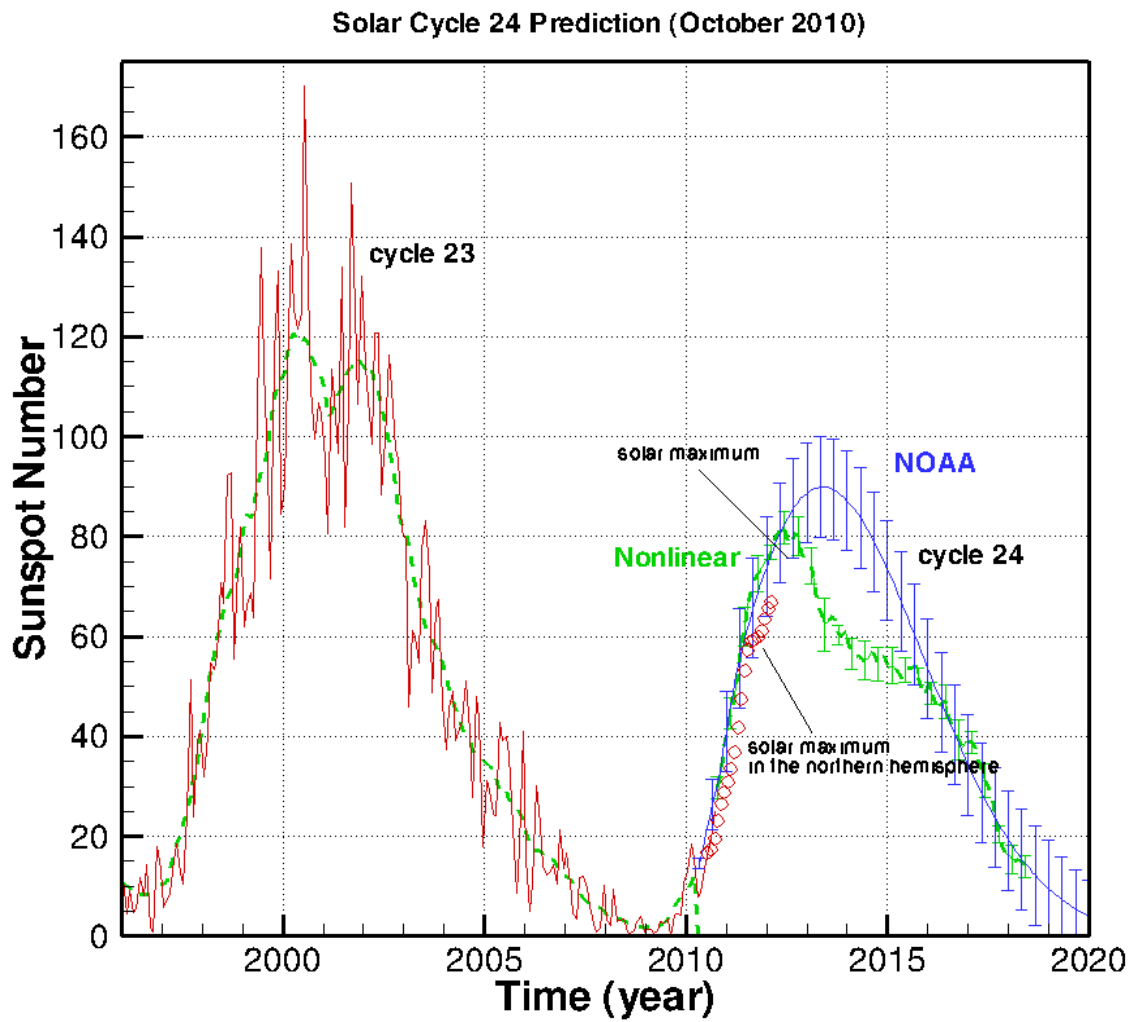

Figure 1- Observed monthly smoothed sunspot numbers (red circles) updated to September 2012, and two predictions (October 2010) from nonlinear method (green line) and from NOAA panel (blue line).

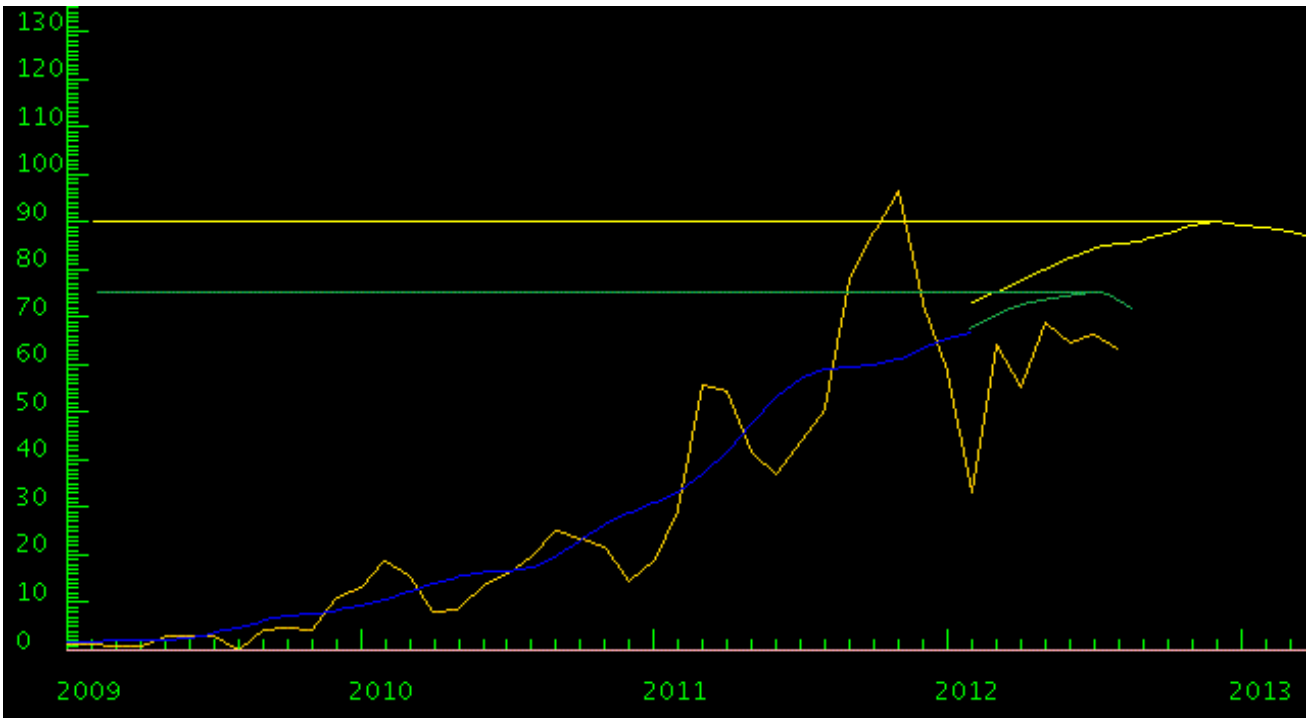

Figure 2- Observed monthly sunspot numbers and monthly smoothed sunspot numbers (blue line) updated to September 2012, with a prediction from IPS (yellow line) reaching a peak of 90.2 at December 2012, compared to our nonlinear prediction (green line) reaching a peak of 75 at June 2012. In order to check the accuracy of the predictions, we need to wait the monthly sunspot data at least up to December 2012 (adapted from IPS).